\documentclass[12pt]{article}
\usepackage{amsmath}
\usepackage{amscd}
\usepackage{amssymb}
\usepackage{array}

\begin{document}
\rightline{CERN-TH/2000-380}
\newcommand{\Z}{\mathbb{Z}}
\newcommand{\R}{\mathbb{R}}
\newcommand{\C}{\mathbb{C}}
\newcommand{\g}{\mathcal{G}}
\newcommand{\s}{\mathcal{S}}
\newcommand{\ka}{\mathcal{K}}
\newcommand{\p}{\mathcal{P}}
\newcommand{\A}{\mathcal{A}}
\newcommand{\I}{\mathcal{I}}
\newcommand{\Ca}{\mathcal{C}}
\newcommand{\N}{\mathbb{N}}
\newcommand{\La}{\mathcal{L}}
\newcommand{\Ha}{\mathbb{H}}

\vskip 2cm

  \centerline{\LARGE \bf On the Embedding of Space-Time }

  \smallskip

\centerline{\LARGE \bf Symmetries into Simple Superalgebras}
\vskip 3cm \centerline{ R. D'Auria$^\dagger$, S. Ferrara$^\star$
and M. A. Lled\'o$^\dagger$.}

\vskip 2cm

\centerline{\it $^\dagger$ Dipartimento di Fisica, Politecnico di
Torino,} \centerline{\it Corso Duca degli Abruzzi 24, I-10129
Torino, Italy, and} \centerline{\it INFN, Sezione di Torino,
Italy.}
 \medskip

\centerline{\it $^\star$ CERN, Theory Division, CH 1211 Geneva 23,
Switzerland, and } \centerline{Laboratori Nazionali di Frascati,
INFN, Italy.}

\vskip 2cm

\begin{abstract}
We explore the embedding of Spin groups of arbitrary dimension and
signature into simple superalgebras in the case of extended
supersymmetry. The R-symmetry, which generically is not compact,
can be chosen compact for all the cases that are congruent mod 8
to the physical conformal algebra so($D-2$,2), $D\geq 3$. An
$\rm{so}(1,1)$ grading of the superalgebra is found in all cases.
Central extensions of super translation algebras are studied in
this framework.


\end{abstract}

\vfill\eject

\section{Introduction}

Supersymmetry algebras in higher dimensions \cite{na,st} have been
the subject of investigation in recent years in connection with
their relation to string and M theories \cite{hw}. The AdS-CFT
correspondence (for a review, see Ref. \cite{agmoo}), which
describes a duality between world-volume brane theories and AdS
supergravities, has renewed the interest in superconformal field
theories and their underlying superalgebras \cite{to, gu, fp, vp}.
It has also been conjectured \cite{hw, to, gu, fp, ba, ho} that
brane superalgebras are closely related to the extension of
superconformal algebras to larger algebras including antisymmetric
tensor generators.

 The most remarkable example is $\rm{osp}(1| 32,\R)$,
  with  32 real spinor charges. By contraction \cite{df} it gives the
M-theory superalgebra (super Poincar\'e in an 11-dimensional space
of signature $(10,1)$), with the maximal central extension (two
and five brane charges). It also can be seen as the super
conformal algebra in a space of dimension  $D=10$ and signature
$(9,1)$ \cite{to,gu}. The conformal group in $D=10$ acts linearly
on a space of dimension $12$, $(10,2)$, and the spinor charge is a
real Weyl spinor in this space. The (1,0) chiral $D=10$ super
Poincar\'e algebra with a five-brane central charge is a
subalgebra of $\rm{osp}(1| 32,\R)$.

Moving to one dimension more, the M-theory super Poincar\'e
algebra (with two and five-brane charges) can be regarded as a
subalgebra of the superconformal algebra  of a space time of
dimension 11. This superalgebra is $\rm{osp}(1| 64,\R)$, whose odd
part (64 spinorial charges) forms a real spinor in a space of
dimension $13$, $(11,2)$ \cite{vv, dflv}. This example is relevant
because it has been shown that M-theory can be regarded as a phase
 of an $\rm{osp}(1| 64,\R)$
theory \cite{we}. The full symmetry is spontaneously broken, so it
is realized non linearly. Some attempts to derive M-theory from a
gauge theory of the group $\rm{Sp}(32,\R)$ have been made in Ref.
\cite{ho}

In this paper we generalize the classification of space-time
superalgebras given in  Ref. \cite{dflv}, in terms of the
dimension $D$ and signature $\rho$ of space time, to extended
supersymmetry algebras with an R-symmetry group acting on the
conformal spinors.

Other than brane superalgebras, the present results are in
relation with different versions of supersymmetric theories with
$D\leq 11$, in particular with the theories introduced by Hull
\cite{hu}, M, M$^*$ and M', which are formulated in 11 dimensional
space-times with signatures $(s,t)=(10,1), (9,2),(6,5)$
respectively. Notice that the signature for all these theories is
$\rho=s-t=\pm1$ mod 8. Our analysis implies that the superalgebras
appearing in them  can be seen  as a contraction of the very same
$\rm{osp}(1| 32,\R)$ algebra or as a subalgebra of $\rm{osp}(1|
64,\R)$. This suggests that they are all different phases of the
same theory \cite{hu, fe}.

The present study evidences  the deep relation between the
R-symmetry algebra and the space-time algebra (the
$\rm{Spin}(s,t)$-algebra which embeds $\rm{so}(s,t)$, according to
the notation of \cite{dflv} ) that appear as simple factors in the
bosonic part of the superconformal algebra. For the signature of
the physical conformal group, $(s,t)=(D-2,2)$ a compact R-symmetry
group is allowed. For the Euclidean superconformal algebra, that
is, for signature $(s,t)=(D-1,1)$ only non compact R-symmetry
groups are allowed. This implies that a proper treatment of
Euclidean supersymmetric theories require superalgebras with non
compact R-symmetry, as observed long ago by Zumino \cite{zu}, and
recently discussed in \cite{si,mc,bvv}.

The paper is organized as follows. In Section 2 we extend the
analysis of ref. \cite{dflv} to N supersymmetries. In Section 3.
we show how, in any dimension, the superconformal algebra has an
so(1,1) grading. The super  Poincar\'e algebra emerges as a non
semisimple subalgebra of the superconformal algebra. In Section 4.
we consider N-extended super Poincar\'e algebras with maximal
central extension.

\section{Conformal superalgebras with $N$ supersymmetries}

We recall here briefly the formalism used in Ref. \cite{dflv}. Let
 $S$ be a complex vector space. A conjugation of $S$ is
an antilinear  map $\sigma:S\rightarrow S$,
$$\sigma(av)=a^*\psi(v), \qquad a\in \C,\; v\in S,$$ such that
$\sigma^2=1$. The set $$S^\sigma=\{v\in S|\sigma(v)=v\}$$ is a
real vector space. If instead $\sigma^2=-1$ we say that $\sigma$
is a pseudoconjugation of $S$, and the condition $\sigma(v)=v$ is
inconsistent.

 Suppose
now that  $S$ is  a $\g$-module, for a complex Lie algebra $\g$.
Let $\sigma:S\rightarrow S$ be an antilinear map satisfying
$\sigma^2=\pm1$ (that is, $\sigma$ is a conjugation or a
pseudoconjugation on $S$). Then the map $$\psi(X)=\sigma\circ
X\circ\sigma^{-1}$$ is a conjugation of $\g$. If $S$ is
irreducible one can prove that  $\g^\psi$ is a real form of $\g$.
The action of $\g^\psi$ on $S$ commutes with $\sigma$.

The real forms of classical Lie algebras arise in this way, except
for the unitary algebras ${\rm su}(p,q)$. The algebras ${\rm
sl}(n,\R)$, ${\rm so}(p,q)$, ${\rm sp}(2p,\R)$ correspond to
conjugations on the space of the fundamental representation and
the algebras ${\rm su}^*(2p)$, ${\rm so}^*(2p)$ and ${\rm
usp}(2p,2q)$ correspond to pseudoconjugations.

\smallskip

Let $V$ be a real vector space of dimension $D=s+t$ with  a non
degenerate symmetric bilinear form  of signature  $\rho=s-t$. We
denote by ${\rm Spin}(s,t)$ the corresponding spin  group. Spinor
representations are linear, orthogonal or symplectic depending on
the existence of a ${\rm Spin}(s,t)$-invariant  bilinear form,
which can be symmetric or antisymmetric. The existence and
symmetry of the bilinear form depend exclusively on the dimension
$D$ modulo 8 \cite{de}. The complexified group ${\rm
Spin}(s,t)^\C$ is then embedded into a classical Lie group, ${\rm
Sl}(n,\C)$, ${\rm SO}(n,\C)$ or ${\rm Sp}(n,\C)$ where $n$ is the
dimension of the spinor representation. Denoting by $D=D_0$ mod(8)
and  $\rho=\rho_0$ mod(8), we have that  for $D_0=2,6$ the spinors
are linear, for $D_0=0, 1,7$ they are orthogonal and for $D_0=3,4,
5$ they are symplectic.

The existence of a conjugation or a pseudoconjugation in the space
$S$ of the spinor representation (commuting with the action of the
orthogonal algebra) depends on the signature $\rho$ mod 8. We have
then that ${\rm o}(s,t)$ is embedded into a real form of the above
algebra, which will be determined in terms of $\rho$. This real
algebra is called the ${\rm Spin}(s,t)$ algebra. In Table
\ref{spinalgebra} the real form for each case is listed
\cite{dflv}. The odd and even cases are treated separately.

\begin{table}[ht]
\begin{center}
\begin{tabular} {|l |l|l|}
\hline Orthogonal&Real, $\rho_0=1,7$&${\rm
so}(2^{\frac{(D-1)}{2}},\R)$ if $D=\rho$\\  \cline{3-3}
$D_0=1,7$&& ${\rm
so}(2^{\frac{(D-1)}{2}-1},2^{\frac{(D-1)}{2}-1})$
 if
$D\neq\rho$\\\cline{2-3} & Quaternionic,\;$\rho_0=3,5$&${\rm
so}^*(2^{\frac{(D-1)}{2}})$\\\hline\hline Symplectic&Real,
$\rho_0=1,7$& ${\rm sp}(2^{\frac{(D-1)}{2}},\R)$\\\cline{2-3}
$D_0=3,5$& Quaternionic,\;$\rho_0=3,5$& ${\rm
usp}(2^{\frac{(D-1)}{2}},\R)$  if $D=\rho$\\\cline{3-3} && ${\rm
usp}(2^{\frac{(D-1)}{2}-1},2^{\frac{(D-1)}{2}-1})$ if $D\neq
\rho$\\\hline\hline\hline
 Orthogonal&Real, $\rho_0=0$&${\rm
so}(2^{\frac{D}{2}-1},\R)$ if $D=\rho$\\  \cline{3-3} $D_0=0$&&
${\rm so}(2^{\frac{D}{2}-2},2^{\frac{D}{2}-2})$
 if
$D\neq\rho$\\\cline{2-3}
 & Quaternionic,\;$\rho_0=4$&${\rm
so}^*(2^{\frac{D}{2}-1})$\\\cline{2-3} &Complex, $\rho_0=2,6$&
${\rm so}(2^{\frac{D}{2}-1},\C)_\R$\\\hline\hline Symplectic&Real,
$\rho_0=0$& ${\rm sp}(2^{\frac{D}{2}-1},\R)$\\\cline{2-3} $D_0=4$&
Quaternionic,\;$\rho_0=4$& ${\rm usp}(2^{\frac{D}{2}-1},\R)$ if
$D=\rho$\\\cline{3-3} && ${\rm
usp}(2^{\frac{D}{2}-2},2^{\frac{D}{2}-2})$ if $D\neq
\rho$\\\cline{2-3} &Complex, $\rho_0=2,6$&${\rm
sp}(2^{\frac{D}{2}-1},\C)_\R$\\\hline\hline

 Linear&Real,
$\rho_0=0$& ${\rm sl}(2^{\frac{D}{2}-1},\R)$\\\cline{2-3}
$D_0=2,6$& Quaternionic,\;$\rho_0=4$& ${\rm
su}^*(2^{\frac{D}{2}-1})$ \\\cline{2-3} &Complex, $\rho_0=2,6$&
${\rm su}(2^{\frac{D}{2}-1})$ if $D= \rho$\\\cline{3-3} &&${\rm
su}(2^{\frac{D}{2}-2},2^{\frac{D}{2}-2})$ if $D\neq \rho$\\\hline
\end{tabular}
\caption{Spin$(s,t)$ algebras.}\label{spinalgebra}
\end{center}
\end{table}

We consider now superalgebras containing ${\rm so}(s,t)$ in the
even part and  whose odd part  consists on one or more copies of
the spinor representation of ${\rm so}(s,t)$. If the superalgebra
is required to be simple, then the anticommutators of the odd
charges generate the whole even subalgebra. In particular, the
generators of the orthogonal group in the spinor representation
must have the appropriate symmetry to appear in the right hand
side of the anticommutator of two spinors. This analysis was
carried out in Ref. \cite{dflv} for $N=1$ supersymmetry. The
smallest simple superalgebra containing ${\rm so}(s,t)$ has even
part equal to the ${\rm Spin}(s,t)$ algebra, listed in Table
\ref{spinalgebra}, times an R-symmetry algebra which is  usp(2) or
so$^*$(2) (quaternionic case) or u(1) (complex
case)\footnote{Except for $D=7$, $\rho=3$, where a smaller simple
superalgebra, the exceptional superalgebra $f(4)$, contains the
orthogonal group in its even part.}. This superalgebra is called
${\rm Spin}(s,t)$ superalgebra. The same structure appears in the
case of extended supersymmetry, although the generalization is not
completely straightforward since new cases appear due to the
presence of the internal index space.

The odd part of the superalgebra is, as a vector space, a tensor
product $S\otimes W$ of the spinor representation space $S$  with
the R-symmetry space $W$. If the spin algebra is a symplectic
algebra, the R-symmetry is an orthogonal algebra and they build an
orthosymplectic algebra. If the spin algebra is orthogonal, then
the R-symmetry is symplectic and they give an orthosymplectic
algebra with the roles of the orthogonal and symplectic groups
interchanged. If the spin algebra is a linear (unitary) algebra,
then the R-symmetry is also a linear (unitary) algebra and they
build a superalgebra form the linear (unitary) series.

In order to obtain a real superalgebra, a conjugation commuting
with the action of the even part of the superalgebra must exist in
the total space $S\otimes W$. If the spinor is real then there is
a conjugation $\sigma_S$ of $S$ commuting with the action of the
Spin($s,t$) algebra.  The R-symmetry factor which acts on $W$ also
commutes with  a conjugation, $\sigma_W$.  If the spinor is
quaternionic then there is a pseudoconjugation $\sigma_S$ on $S$,
and  there is also a pseudoconjugation on $W$ commuting with the
R-symmetry. Then $\sigma_S\otimes\sigma_W$ is a conjugation in the
total space. If the spinor is complex,  the Spin($V$) algebra is
either ${\rm su}(p,q)$ or a  complex group (symplectic or
orthogonal). In the first case the R-symmetry is also ${\rm
su}(m,n)$
and in the second case the R-symmetry is a complex group
(orthogonal or symplectic respectively). The real representation
is obtained by taking the complex vector space as a real one of
twice the dimension.

As an example, we compute the case $D_0=1,7$, $\rho_0=1,7$. The
spinors are orthogonal and real. The anticommutator of two odd
generators is of the form
\begin{equation}\{Q_\alpha^i,Q_\beta^j\}=\sum_k A^{ij}
\gamma_{\alpha\beta}^{[\mu_1\cdots \mu_k]}Z_{[\mu_1\cdots
\mu_k]}.\label{antico}\end{equation}  The generators of the
orthogonal group are $Z_{[\mu_1 \mu_2]}$. For $N=1$ the factor
$A^{ij}$ is not present. Then, in order to have $Z_{[\mu_1
\mu_2]}$ in the right hand side of (\ref{antico}), the morphism
$\gamma^{[\mu_1 \mu_2]}_{\alpha\beta}$ must be symmetric. Since it
is antisymmetric, there is no superconformal algebra in this case.
For $N>1$, one can choose an antisymmetric matrix
$A^{ij}=\epsilon^{ij}$ and the orthogonal generators are allowed
in the right hand side of (\ref{antico}). If follows that the
R-symmetry group is ${\rm Sp}(2N,\R)$.

In Table \ref{next} we list the R-symmetry groups and the ${\rm
Spin}(s,t)$ superalgebras. The compact cases $D=\rho$ are not
listed but they are immediate. We mark with the symbol ``$\circ$"
the cases that do not arise in the  non extended case.

The cases  marked with a symbol ``$\star$" allow the possibility
of a compact R-symmetry group. They correspond to ${\rm so}(s,2)$,
that is the physical conformal groups.

\begin{table}[ht]
\begin{center}
\begin{tabular} {c|c|c|l |l|}

\cline{2-5} &$D_0$&$\rho_0$& R-symmetry&Spin$(s,t)$
superalgebra\\\cline{2-5} $\circ$& 1,7& 1,7& ${\rm
sp}(2N,\R)$&${\rm
osp}(2^{\frac{D-3}{2}},2^{\frac{D-3}{2}}|2N,\R)$\\\cline{2-5}
$\star$& 1,7& 3,5& ${\rm usp}(2N-2q,2q)$&${\rm
osp}(2^{\frac{D-1}{2}\,*}|2N-2q,2q)$\\\cline{2-5} $\star$& 3,5&
1,7& ${\rm so}(N-q,q)$&${\rm
osp}(N-q,q|2^{\frac{D-1}{2}})$\\\cline{2-5} & 3,5& 3,5& ${\rm
so}^*(2N)$&${\rm
osp}(2{N}^*|2^{\frac{D-3}{2}},2^{\frac{D-3}{2}})$\\\cline{2-5}
\cline{2-5} $\circ$& 0& 0& ${\rm sp}(2N,\R)$&${\rm
osp}(2^{\frac{D-4}{2}},2^{\frac{D-4}{2}}|2N)$\\\cline{2-5}
$\circ$& 0& 2,6& ${\rm sp}(2N,\C)_\R$&${\rm
osp}(2^{\frac{D-2}{2}}|2N,\C)_\R$\\\cline{2-5} $\star$& 0& 4&
${\rm usp}(2N-2q,2q)$&${\rm
osp}(2^{\frac{D-2}{2}\,*}|2N-2q,2q)$\\\cline{2-5} & 2,6& 0& ${\rm
sl}(N,\R)$&${\rm sl}(2^{\frac{D-2}{2}}|N,\R)$\\\cline{2-5}
$\star$&2,6& 2,6& ${\rm su}(N-q,q)$&${\rm
su}(2^{\frac{D-4}{2}},2^{\frac{D-4}{2}}|N-q,q)$\\\cline{2-5}
$\circ$& 2,6& 4& ${\rm su}^*(2N,\R)$&${\rm
su}(2^{\frac{D-2}{2}}|2N)^*$\\\cline{2-5} $\star$& 4& 0& ${\rm
so}(N-q,q)$&${\rm osp}(N-q,q|2^{\frac{D-2}{2}})$\\\cline{2-5} & 4&
2,6& ${\rm so}(N,\C)_\R$&${\rm
osp}(N|2^{\frac{D-2}{2}},\C)_\R$\\\cline{2-5} & 4& 4& ${\rm
so}^*(2N)$&${\rm
osp}(2{N}^*|2^{\frac{D-4}{2}},2^{\frac{D-4}{2}})$\\\cline{2-5}
\end{tabular}
\caption{Spin$(s,t)$ superalgebras.}\label{next}
\end{center}
\end{table}

\section{${\rm so}(1,1)$ grading of the ${\rm Spin}(s,t)$ superalgebra}

Let $\g^k$ be a compact semisimple Lie algebra, its
complexification being $\g^c$. Let $\theta:\g^k\mapsto \g^k$ be an
involutive automorphism, $\theta^2=1$. $\g^k$ splits into two
eigenspaces, $$\g^k=\ka+\p.$$ $\ka$ is the eigenspace with
eigenvalue $+1$ of $\theta$, and $\p$ the eigenspace with
eigenvalue $-1$. The vector space
\begin{equation}\g=\La_0+i\p\label{cd}\end{equation} is a non
compact real form of $\g^c$. (\ref{cd}) is called a {\it Cartan
decomposition} of $\g$ and the procedure is known as the Weyl
unitary trick. All the Cartan decompositions are listed in
Ref.\cite{he}. $\La_0$ is the maximal compactly embedded
subalgebra of $\g$, and $\p$ carries an irreducible representation
of $\La_0$.

If $\g$ is simple, then the algebra $\ka$ is either semisimple or
is a semisimple algebra plus  a u(1) factor. For example, let
$\g^k=\rm{so}(p+q)$ and $\theta_{p,q}(X)=I_{p,q}XI_{p,q}$, where
$$I_{p,q}=\begin{pmatrix}I_p&0\\0& -I_q\end{pmatrix}.$$ Then
$\ka={\rm so}(p)\oplus \rm{so}(q)$. $\p$ is the bifundamental
representation of ${\rm so}(p)\oplus \rm{so}(q)$,
$(\mathbf{p,q})$ .

One can apply  the Weyl unitary trick  to $\g$ with respect to a
Cartan involution $\theta'$ that commutes with $\theta$. $\g$
splits into four eigenspaces of the simultaneous eigenvalues of
$\theta$ and $\theta'$, $$\g=\g^{++}+\g^{+-}+i\g^{-+}+i\g^{--}.$$
The first sign is the eigenvalue of $\theta$ and the second is the
eigenvalue of $\theta'$. So $$\ka=\g^{++}+\g^{+-}, \qquad
\p=\g^{-+}+\g^{--}.$$ The Lie algebra
$$\g'=\g^{++}+i\g^{+-}+i\g^{-+}+\g^{--}$$ is another non compact
form of $\g^c$ (corresponding to the Cartan involution
$\theta'\circ\theta$). The maximal compact subalgebra is
$\g^{++}+\g^{--}$. We take the decomposition
$$\La_0=\g^{++}+i\g^{+-},\qquad \p'=i\g^{-+}+\g^{--}.$$ $\La_0$ is
a non compact real form of the complexification of $\ka$.

In the  example  of $\rm{so}(d)$, we can take
$\theta=\theta_{d-2,2}$ and $\theta'=\theta_{s,t}$, $s+t=d$. Then
$$\g'=\rm{so}(s,t), \qquad
\La_0=\rm{so}(s-1,t-1)\oplus\rm{so}(1,1)$$ and the bifundamental
 representation $(\mathbf{d-2,2})$ splits into two irreducible
 representations, whith charges $\pm1$ with respect to
 $\rm{so}(1,1)$, $(\mathbf{d-2})^{+1}\oplus(\mathbf{d-2})^{-1}$. This is in fact
 the splitting of the conformal algebra with respect to the
 Lorentz subalgebra times the dilatation. The vector space $\p'$
 contains the translations $P_\mu$ and the conformal boosts
 $K_\mu$.

 The Lie algebra $\g'=\rm{so}(s,t)$ has a Lie algebra grading,
$$\g'=\La_{-1}+\La_0+\La_{+1},$$
 being the grade the charge with respect to $\rm{so}(1,1)$.
 $\La_{\pm 1}$ are abelian subalgebras and $\La_{0} \circledS\La_{\pm
 1}$ is the Poincar\'e algebra.

 The same procedure can be applied to all the Spin$(s,t)$ algebras
 in  Table \ref{spinalgebra}, and a grading is always found with respect
 to a $\rm{so}(1,1)$ subalgebra, which can always be identified
 with the dilatation of the conformal group embedded into the
 Spin$(s,t)$ algebra. $\La_0\circledS\La_{\pm
 1}$ is a semidirect sum of algebras that generalizes the super Poincar\'e
 algebra. $\La_{\pm 1}$ are abelian algebras containing the translations and central charges.
  These decompositions are listed in \cite{gi}
 for all the simple Lie algebras. In Table \ref{grade} we give
 $\La_0$ for all the Spin$(s,t)$ algebras.

\begin{table}[ht]
\begin{center}
\begin{tabular} {|c|c|l|}

\hline

Spin$(s,t)$ algebra& $\La_0$&Fundamental
Representation\\\hline\hline

 $\rm{so}(n,n)$&$\rm{sl}(n,\R)\oplus\rm{so}(1,1)$&$\mathbf{2n}=(\mathbf{n})^{1/2}\oplus
 (\mathbf{n'})^{-1/2}$

\\\hline

$\rm{so}^*(4n)$&$\rm{su}^*(2n)\oplus\rm{so}(1,1)$&$\mathbf{4n}=(\mathbf{2n})^{1/2}\oplus
 (\mathbf{2n'})^{-1/2}$

\\\hline

$\rm{so}(2n,\C)$&$\rm{gl}(n,\C)$&$\mathbf{2n}=(\mathbf{n})^{1/2}\oplus
 (\mathbf{n'})^{-1/2}$

\\\hline

$\rm{sp}(2n,\R)$&$\rm{sl}(n,\R)\oplus\rm{so}(1,1)$&$\mathbf{2n}=(\mathbf{n})^{1/2}\oplus
 (\mathbf{n'})^{-1/2}$

\\\hline

$\rm{usp}(2n,2n)$&$\rm{su}^*(2n)\oplus\rm{so}(1,1)$&$\mathbf{4n}=(\mathbf{2n})^{1/2}\oplus
 (\mathbf{2n'})^{-1/2}$

\\\hline

$\rm{sp}(2n,\C)$&$\rm{gl}(2n,\C)$&$\mathbf{2n}=(\mathbf{n})^{1/2}\oplus
 (\mathbf{n'})^{-1/2}$

\\\hline

$\rm{sl}(2n,\R)$&$\rm{sl}(n,\R)\oplus\rm{sl}(n,\R)\oplus
$&$\mathbf{2n}=(\mathbf{n},1)^{1/2}\oplus
 (1,\mathbf{n'})^{-1/2}$\\
 &\rm{so}(1,1)&$\mathbf{2n'}=(\mathbf{n'},1)^{-1/2}\oplus
 (1,\mathbf{n})^{1/2}$

\\\hline

$\rm{su}^*(4n)$&$\rm{su}^*(2n)\oplus\rm{su}^*(2n)\oplus
$&$\mathbf{4n}=(\mathbf{2n},1)^{1/2}\oplus
 (1,\mathbf{2n'})^{-1/2}$\\
 &\rm{so}(1,1)&$\mathbf{4n'}=(\mathbf{2n'},1)^{-1/2}\oplus
 (1,\mathbf{2n})^{1/2}$

\\\hline

$\rm{su}(n,n)$&$\rm{sl}(n,\C)\oplus
$&$\mathbf{2n}=(\mathbf{n})^{1/2}\oplus
 (\mathbf{{\bar n}'})^{-1/2}$\\
 &\rm{so}(1,1)&$\mathbf{2{\bar n}}=(\mathbf{{\bar n}})^{1/2}\oplus
 (\mathbf{n'})^{-1/2}$

\\\hline

\end{tabular}
\caption{$\rm{so}(1,1)$-grading.}\label{grade}
\end{center}
\end{table}

When the Spin$(s,t)$ algebra is an orthogonal algebra,
$\La_{\pm1}$ are in the two-fold antisymmetric  representation of
$\La_0$. When the Spin$(s,t)$ algebra is a symplectic algebra,
then $\La_{\pm1}$ are in the two-fold symmetric representation of
$\La_0$. When the Spin$(s,t)$ algebra is a linear algebra, then
$\La_{\pm1}$ are in the bifundamental  representation of $\La_0$.

\smallskip

The spinor representation of ${\rm so}(s,t)$, $S_{(s,t)}$
decomposes as \begin{equation}
\begin{CD}S_{(s,t)}@>>{\rm so}(s-1,t-1)\oplus{\rm
so}(1,1)>S_{(s-1,t-1)}^{1/2}\oplus
S_{(s-1,t-1)}^{-1/2}\end{CD}.\label{split}\end{equation}  If $D$
is even, then a chiral representation decomposes into two
representations with opposite chirality.

$\rm{so}(s-1,t-1)$ is embedded into  $\La_0$. When promoting the
spinor representation of $\rm{so}(s,t)$ to the fundamental
representation of the Spin$(s,t)$ algebra, the splitting
(\ref{split}) corresponds to  the  splitting under $\La_0$, which
is given also in Table \ref{grade}. $\mathbf{n'}$ denotes the dual
representation of $\mathbf{n}$.

In fact, one can check that $\La_0$ contains in each case, the
full Spin$(s-1,t-1)$-algebra. The embeddings are given in Table
\ref{emb}.

\begin{table}[ht]
\begin{center}
\begin{tabular} {|c||c||c|}

\hline Real case& Quaternionic case&Complex case\\\hline\hline

 $\rm{sl}(2n,\R)\supset\rm{so}(n,n)$
 &$\rm{su}^*(2n)\supset\rm{so}^*(2n)$&$\rm{sl}(2n,\C)\supset\rm{so}(2n,\C)$

\\\hline

$\rm{sl}(2n,\R)\supset\rm{sp}(2n,\R)$&$\rm{su}^*(2n)\supset\rm{usp}
(n,n)$&$\rm{sl}(2n,\C)\supset\rm{sp}(2n,\C)$

\\\hline

\end{tabular}
\caption{Embedding of Spin$(s,t)$ algebras.}\label{emb}
\end{center}
\end{table}
The superalgebras of Table \ref{next} have also a $\rm{so}(1,1)$
grading $$\s=\La_{-1}+\s_{-1/2}+\La_{0}+\s_{+1/2}+\La_{+1}.$$ In
the case of extended supersymmetry, $\La_0$ contains also the
R-symmetry factor. The spaces $\La^{\pm1}$ have a simple meaning
in terms of the $\gamma$-matrices of $\rm{so}(s-1,t-1)$. From the
grading properties and the simplicity of the superalgebra,  it is
clear that $$\{\s_{\pm1/2},\s_{\pm1/2}\}=\La_{\pm1}.$$ In fact,
$\La_\pm$ are irreducible representations of $\La_0$. The
anticommutator $$\{Q_{\pm1/2},Q_{\pm1/2}\}$$ can in general be
written as in (\ref{antico}), and the term $\gamma^\mu$
($\mu=1,\dots, s+t-2$) appears since it corresponds to the
momentum (with grade 1). It follows that in the r.h.s of
(\ref{antico}) will appear only terms corresponding to
 matrices $\gamma^{[\mu_1\dots \mu_m]}$ with the same symmetry properties as
$\gamma^\mu$. In fact, from dimensional considerations, all of
these terms appear. For $s+t=7,8,9$ mod 8, the  $\gamma^\mu$ are
antisymmetric and  dim$(\La_\pm)=\frac{n(n-1)}{2}$. For
$s+t=3,4,5$ mod 8, they are symmetric and
dim$(\La_\pm)=\frac{n(n+1)}{2}$. For $s+t=2,6$ mod 8, the relevant
anticommutator is  a left spinor with a right spinor. In this case
the morphisms (coefficients in the r.h.s. of (\ref{antico})) have
no definite symmetry  and dim$(\La_\pm)=n^2$.

\section{The orthosymplectic algebra and the  maximal central extension
of  Poincar\'e supersymmetry }

Let $2n$ be the real dimension of a spinor representation and $N$
the number of such spinors present in the superalgebra. The spinor
charges are denoted $$Q^i_\alpha, \qquad i=1,\dots N,
\;\;\;\alpha=1,\dots 2n.$$ Since the anticommutator
$\{Q^i_\alpha,Q^j_\beta\}$ is symmetric, the biggest simple
superalgebra containing only these odd generators (maximal
Spin$(s,t)$ algebra in the language of Ref. \cite{dflv}) is
$\rm{osp}(1|2nN, \R)$, with  bosonic part
$\hat\La=\rm{sp}(2nN,\R)$. It is clear that $\La$ contains as a
subalgebra the Spin$(s,t)$ algebra plus the R-symmetry,

$$ \La \oplus \rm{R}\!\!-\!\!\rm{symmetry}\,\subset \hat\La.$$
 We
want to show that the so(1,1) grading of $\La$  extends to
$\hat\La$ and to $\rm{osp}(1|2nN)$.  We have that $$
\hat\La=\hat\La_{+1}+\hat\La_0+\hat\La_{-1},$$ where
$\hat\La_0=\rm{sl}(nN,\R)\oplus \rm{so}(1,1)$ and $\hat\La_{\pm1}$
are in the two-fold symmetric representation of $\rm{sl}(nN,\R)$
with charges $\pm1$ with respect to $\rm{so}(1,1)$. To show that
this grading is compatible with the one of the  Spin$(s,t)$
algebra, we have to show that $$ \La_0 \oplus
\rm{R}\!\!-\!\!\rm{symmetry}\subset \hat\La_0.$$

We consider the complex linear algebra
$$\rm{sl}(nN,\C)\simeq\rm{gl}(n,\C)\otimes\rm{gl}(N,\C)/\C^*.$$
(The bracket in the tensor product of algebras is defined as
$$[a\otimes a',b\otimes b']=[a,b]\otimes[a',b']).$$ One has that
\begin{eqnarray*}&\rm{sl}(n,\C)\simeq\rm{sl}(n,\C)\otimes
\rm{Id}\subset\rm{sl}(nN,\C)\\ & \rm{sl}(N,\C)\simeq
\rm{Id}\otimes\rm{sl}(N,\C)\subset\rm{sl}(nN,\C),\end{eqnarray*}
and that
$$\rm{sl}(n,\C)\oplus\rm{sl}(N,\C)\subset\rm{sl}(nN,\C).$$ Notice
that the fundamental representation of $\rm{sl}(nN,\C)$ becomes
the bifundamental ($\C^n\otimes\C^N$) of
$\rm{sl}(n,\C)\oplus\rm{sl}(N,\C)$. The adjoint of
$\rm{sl}(nN,\C)$ decomposes under
$\rm{sl}(n,\C)\oplus\rm{sl}(N,\C)$ in the adjoint of
$\rm{sl}(n,\C)\oplus\rm{sl}(N,\C)$ plus the tensor products of the
adjoints of  $\rm{sl}(n,\C)$ and  $\rm{sl}(N,\C)$.

Let $\g_n\subset\rm{sl}(n,\C)$ and $\g_N\subset\rm{sl}(N,\C)$.
Then we have $$\g_n\oplus \g_N\subset \rm{sl}(nN,\C).$$ In
particular, $\g_n$ and $\g_N$ can be orthogonal, symplectic, or
the linear algebras themselves. We have just to check that the
appropriate real forms are contained in $\rm{sl}(nN,\R)$. We
recall that the real form $\rm{sl}(nN,\R)$ is obtained from a
conjugation in the fundamental representation space. Moreover,
$\rm{sl}(nN,\R)$ is the set of all traceless matrices in
$\rm{sl}(nN,\C)$, commuting with a conjugation in $\C^{nN}$ (which
can be brought by an isomorphism to the standard conjugation on
$\C^{nN}$).

The direct sum of  real forms $\g^r_{n}\oplus \g^r_{N}$ is
contained in $\rm{sl}(nN,\R)$ if the bifundamental representation
of $\g^r_{n}\oplus \g^r_{N}$ commutes with a conjugation. This
happens when both, $\g^r_{n}$ and $\g^r_{N}$, commute with a
conjugation in their respective spaces, $\C^n$ and $\C^N$, and
when they both commute with a pseudoconjugation.

If the algebras are complex, then one can see  $(\g_{n'})_\R$ and
$(\g_{N'})_\R$ inside a real group
\begin{eqnarray}(\g_{n'})_\R\subset
\rm{sl}(n',\C)_\R\subset\rm{sl}(2n',\R))\subset
\rm{sp}(4n',\R),\nonumber\\ (\g_{N'})_\R\subset
\rm{sl}(N',\C)_\R\subset\rm{sl}(2N',\R))\subset
\rm{sp}(4N',\R)\label{real}\end{eqnarray}
 with $n=2n'$, $N=2N'$.

The unitary algebras are embedded into complex algebras that one
can see as real, and then embedded in a linear complex group as
above (\ref{real}).

We give the embeddings for all dimensions and signatures in Table
\ref{embeddings}. The real dimension of the spinor representation.
 is given in terms of $D$ and depends on the reality
properties of the spinor. For clarity, it is given in Table
\ref{dimensions}.

The grading extends to the orthosymplectic superalgebra
$\hat\s=\rm{osp}(1|2m,\R)$, where $m$ is the number appearing in
the third column of Table \ref{embeddings},
$$\hat\s=\hat\La_{-1}+\hat\s_{-1/2}+\hat\La_{0}+\hat\s_{+1/2}+\hat\La_{+1}.$$

\begin{table}[ht]
\begin{center}
\begin{tabular} {|c |c|c|c|}
\hline$D_0$&$\rho_0$ &$\hat\La_0$&$\La_0\oplus
\rm{R\!\!-\!\!symmetry} $\\\hline 1,7&1,7
&$\rm{sl}(2N2^\frac{D-3}{2},\R)$&$\rm{sl}(2^\frac{D-3}{2},\R)\oplus\rm{sp}(2N,\R)$\\
\hline
0&0&$\rm{sl}(2N2^\frac{D-4}{2},\R)$&$\rm{sl}(2^\frac{D-4}{2},\R)\oplus\rm{sp}(2N,\R)$\\\hline
1,7&3,5&$\rm{sl}(2N2^\frac{D-3}{2},\R)$&$\rm{su}^*(2^\frac{D-3}{2},\R)\oplus\rm{usp}(2N-2q,2q)$\\\hline

3,5&1,7&$\rm{sl}(N2^\frac{D-3}{2},\R)$&$\rm{sl}(2^\frac{D-3}{2},\R)\oplus\rm{so}(N-q,q)$\\
\hline 4&0&
$\rm{sl}(N2^\frac{D-4}{2},\R)$&$\rm{sl}(2^\frac{D-4}{2},\R)\oplus\rm{so}(N-q,q)$\\\hline

3,5&3,5&$\rm{sl}(2N2^\frac{D-3}{2},\R)$&$\rm{su}^*(2^\frac{D-3}{2})\oplus\rm{so}^*(2N)$\\
\hline
4&4&$\rm{sl}(2N2^\frac{D-4}{2},\R)$&$\rm{su}^*(2^\frac{D-4}{2})\oplus\rm{so}^*(2N)$\\\hline

0&2,6&$\rm{sl}(2N2^\frac{D-2}{2},\R)\supset\rm{sl}(2N2^\frac{D-4}{2},\C)
$&$\rm{sl}(2^\frac{D-4}{2},\C)\oplus\rm{sp}(2N,\C)$\\\hline

2,6&0&$\rm{sl}(N2^\frac{D-2}{2},\R)$&$\rm{sl}(2^\frac{D-4}{2},\R)\oplus
\rm{sl}(2^\frac{D-4}{2},\R)\oplus\rm{sl}(N,\R)$\\\hline

2,6&2,6&$\rm{sl}(N2^\frac{D-2}{2},\R)\supset\rm{sl}(N2^\frac{D-4}{2},\C)
$&$\rm{sl}(N2^\frac{D-4}{2},\C)\oplus\rm{su}(N-q,q)$\\\hline

2,6&4&$\rm{sl}(2N2^\frac{D-2}{2},\R)$&$\rm{su}^*(2^\frac{D-4}{2})\oplus
\rm{su}^*(2^\frac{D-4}{2})\oplus\rm{su}^*(2N)$\\\hline

4&2,6&$\rm{sl}(N2^\frac{D-2}{2},\R)\supset\rm{sl}(N2^\frac{D-4}{2},\C)
$&$\rm{sl}(2^\frac{D-4}{2},\C)\oplus\rm{so}(N,\C)$\\\hline

0&4&$\rm{sl}(2N2^\frac{D-4}{2},\R)$&$\rm{su}^*(2^\frac{D-4}{2},\C)\oplus\rm{usp}(2N-2q,2q)$\\\hline

\end{tabular}
\caption{Graded embeddings} \label{embeddings}
\end{center}
\end{table}

\begin{table}[hb]
\begin{center}
\begin{tabular} {|c |c| c||c|c|c|}
\hline $\rho_0$(odd) &real dim($S$) & reality &$\rho_0$(even)
&real dim($S^{\pm}$) & reality\\ \hline \hline 1& $2^{(D-1)/2}$
&$\R$ & 0 &$2^{D/2-1}$&$\R$ \\ \hline 3& $2^{(D+1)/2}$ &$\Ha$& 2
&$2^{D/2}$&$\C$ \\ \hline 5& $2^{(D+1)/2}$ &$\Ha$& 4
&$2^{D/2}$&$\Ha$\\ \hline 7& $2^{(D-1)/2}$ &$\R$ & 6
&$2^{D/2}$&$\C$ \\ \hline
\end{tabular}
\caption{Real dimensions of spinor
representations}\label{dimensions}
\end{center}
\end{table}
\vfill\eject

To see this, it is enough to give the decomposition of the
fundamental representation of $\rm{sp}(2m, \R)$ with respect to
$\rm{sl}(m,\R)\oplus\rm{so}(1,1)$,
\begin{equation*}
\begin{CD}({\bf 2m})@>>\rm{sl}(m,\R)\oplus\rm{so}(1,1)>({\bf m})^{1/2}\oplus
({\bf m}')^{-1/2}\end{CD}.\end{equation*}

Finally, $\hat\s_{+1/2}+\hat\La_{+1}$ is a superalgebra which is
the maximal central extension of the supertranslation algebra.

\section*{Acknowledgements}

S. F. wants to thank the INFN, Sezione di Torino for its kind
hospitality during the completion of this work.  The work of S. F.
has been supported in part by the European Commission RTN network
HPRN-CT-2000-00131, (Laboratori Nazionali di Frascati, INFN) and
by the D.O.E. grant DE-FG03-91ER40662, Task C.

\end{document}